\documentstyle[prl,epsfig,aps]{revtex}
\begin{document}
\preprint{LPTENS-96/31 Saclay 96}
\twocolumn[\hsize\textwidth\columnwidth\hsize\csname@twocolumnfalse\endcsname
\title{ Glassy behaviour in disordered systems with non-relaxational dynamics}
\author{Leticia F. Cugliandolo{$^*$}, 
Jorge Kurchan{$^{**}$}, Pierre Le Doussal{$^{**}$} and Luca Peliti{$^{***}$}}
\address{ 
$^*$
CNRS-Laboratoire de Physique Th\'{e}orique de l'Ecole Normale 
Sup\'{e}rieure\cite{add1} \\
24, rue Lhomond, F-75231 Paris Cedex 05, France.
\\
$^{**}$ SPEC, Saclay,
Orme des Merisiers, F-91191 Gif--sur--Yvette, France\cite{add2}.
\\
$^{***}$ Laboratoire de Physico-Chimie Th\'eorique\cite{add3}
ESPCI, 10 rue Vauquelin, F-75231 Paris Cedex 05, France 
and \\
Dipartimento di Scienze Fisiche, Unit\`a INFM, \\
Universit\`a ``Federico II'', Mostra d'Oltremare, Pad.~19,
I-80125 Napoli, Italy.
}
\pacs{74.60.Ge, 05.20.-y}
\maketitle
\begin{abstract}
We show that a family of disordered systems with non-relaxational
dynamics may exhibit ``glassy'' behavior at nonzero temperature,
although such a behavior appears to be ruled out by a face-value
application of mean-field theory. Nevertheless, the roots of this
behavior can be understood within mean-field theory itself, properly
interpreted. Finite systems belonging to this family have a dynamical regime
with a self-similar pattern of alternating periods of fast motion
and trapping.
\end{abstract}    

\twocolumn 
\vskip.5pc] 
\narrowtext

The dynamics of disordered physical systems exhibits 
``glassy'' features such as ergodicity breaking, slow dynamics 
 and aging.
These systems are usually modelled by  purely relaxational 
stochastic processes satisfying detailed
balance.
In this situation, glassiness arises from the complexity 
of the energy landscape: the 
representative point strives to go downhill in energy
along the maximal slope while receiving  random kicks from the thermal noise,
and gets trapped into deeper and deeper energy valleys.

Disordered physical systems have often been 
considered a paradigm for complex behaviour in 
other fields, especially biology.
However, there are no compelling reasons, in these fields, to restrict oneself to 
purely relaxational dynamics. It is therefore important to 
know if the glassy properties exhibited by
purely relaxational dynamical models are also present in the general case
and, if so, whether the mechanisms responsible
 for glassy dynamics in non-relaxational
systems~\cite{Crfavu} are completely different from those 
acting in purely relaxational ones, even when the violation
of detailed balance is small. 
If they  were different, the   picture that has evolved
to explain the behaviour of disordered
 systems would be entirely irrelevant as soon as non-dissipative forces are turned on.

Hertz et~al., Parisi ~\cite{Grheso} and 
especially Crisanti and Sompolinsky~(CS)~\cite{CS} 
showed some years ago that spin-glass behaviour was 
destroyed in several mean-field disordered
models by an arbitrarily small but generic violation of 
detailed balance.

We show in this paper that 
there is a wide class of disordered systems
in which glassy behavior 
 resists   non-relaxational
perturbations  (even at $T>0$), provided that either the initial condition is 
properly chosen, or
that the size of the system is finite. 
(We shall not deal here with effects that are exclusive
of the zero temperature case.)
This behavior should be relevant for
infinite systems in a finite dimensional space.
The reason for this robustness can be understood, 
at least qualitatively, by a suitable interpretation of mean-field theory as follows.

 The stability of each energy (resp. free-energy)  saddle
can be characterized by the lowest eigenvalue $\lambda_{\rm min}$
of the Hessian. A class of models, of
which that of Sherrington and Kirkpatrick (SK) is the best-known representative,
are {\em marginal} in the sense that in almost all  states $\lambda_{\rm min}$ goes to zero
in the thermodynamical limit, and the single-state spin-glass susceptibility diverges.
 \cite{Brmo,Neta,Mepavi}.
 It is then not surprising that in such purely marginal models a small (but still
${\rm O}(N)$) perturbation may completely change 
their dynamics, as found in \cite{Grheso,CS}.

However, there is another class of models (currently thought to mimic `fragile' structural glasses)
having many non-marginal states with non-zero  $\lambda_{\rm min}$ and  for which the
spin-glass susceptibility within such states is finite. Non-marginal states look locally
much like ferromagnetic states or the retrieval states
in the Hopfield  model. Now, it is easy to convince oneself
that the stability properties of a   
non-marginal state cannot be dramatically altered by the combined effects
of arbitrarily small non-relaxational forces and thermal noise \cite{other}.

Consider for definiteness the typical case of the $p$-spin spherical spin-glass \cite{Crso}.
The energy landscape of the system~\cite{Kupavi} features many saddle points of the 
energy  function at different energy-density  values   ${\cal E}<0$.
One can identify a threshold value ${\cal E}_{\rm th}$
such that only the saddle points
with ${\cal E}<{\cal E}_{\rm th}$ are minima.
 The lowest eigenvalue $\lambda_{\rm min}$  at each saddle point 
is proportional to the depth of the state beneath  the threshold
${\cal E}_{\rm th}-{\cal E}$.
The only marginal saddle points are those just below the threshold,
unlike the case of marginal models for which all states are marginal 
(or, in other words, for which there are no minima deep below the threshold).
 From the reasoning above, we might expect that  only
near-threshold minima will be destabilized by infinitesimal asymmetries,
whereas the deeper a state, the more robust it will be.

In the $p$-spin spherical glass, purely relaxational dynamics 
 starting from a random initial condition  
exhibits ``aging''
 \cite{Cuku}: when $N=\infty$, the system keeps touring a region just 
above the threshold, moving
slower and slower but without ever getting completely 
trapped. For large but finite $N$ (or for an infinite
system in {\em any\/} finite dimensionality) the system penetrates
a time-dependent amount below the threshold, and still ages due to the increasing
depth of the traps it finds.
In this purely relaxational case, the dynamics is qualitatively 
similar to that of a marginal model.

Now, 
just as a non-relaxational perturbation destroys aging in a marginal model, it also seems
to destroy aging {\em around the threshold\/} in a non-marginal one. 
However, in non-marginal models,
there are an infinite number of deeper states
that remain stable in the presence of the perturbation.
Hence, as soon as a finite system (or an infinite system in finite dimensions) 
is able to
 penetrate below the threshold, it rediscovers the glassy features 
which had been destroyed above and near the threshold.

In what follows we confirm this scenario for non-marginal
models  with non-relaxational dynamics.
We first give evidence that mean-field dynamics ($N=\infty$), starting from a 
random initial condition, yields for long times a time-translational invariant 
solution for the correlation and response functions 
even at small asymmetries (no aging), and that 
the correlation functions decay to zero (no ergodicity breaking), confirming CS~\cite{CS}.

We then show, always within mean-field dynamics,  that there are initial
conditions such that the correlations do not decay to zero,
in the presence of non-relaxational perturbations, even at $T>0$.
This confirms the existence of stable regions with ergodicity breaking.
Obviously, such regions  are never found if the system 
is infinite and starts from a random configuration.

Finally, in order to estimate the time scales involved for trapping in a large but finite system,
we performed simulations.

We consider a system of $N$ variables ${\bf s}=(s_1,\ldots,s_N)$, subject
to forces $F_i$ given by
\begin{equation}
F_i ({\bf s})= \sum_{\{j_1,\ldots,j_{p-1}\}}
 J_{i}^{j_1\ldots j_{p-1}} s_{j_1} \ldots s_{ j_{p-1}},
\end{equation}
where the couplings are random Gaussian variables. For different sets of indices
 $\{i ,j_1,\ldots,j_{p-1}\}$ the $J$'s are uncorrelated, 
while for permutations of the same set
of indices they are correlated so that
\begin{equation}
\overline{F_i({\bf s'})F_j ({\bf s})}= \delta_{ij} f_1(q)+s'_i s_j f_2(q)/N,
\end{equation}
where  $q=({\bf s}\cdot{\bf s}')/N$.  In the purely relaxational model
one has $f_2(q)=f'_1(q)$. We consider here
$f_2(q)=\alpha f_1'(q)$, where $f_1(q)=p  q^{p-1}/2$. The purely
relaxational case has symmetric couplings $ J_{i}^{j_1\ldots j_{p-1}}$
under  $i \leftrightarrow j_k$ ($\alpha=1$)
while for uncorrelated $ J_{i}^{j_1\ldots j_{p-1}}$
and $ J_{j}^{i j_2\ldots j_{p-1}}$: $\alpha=0$.

We consider: (i) the continuous spherical
 model $|{\bf s}|^2=N$, with Langevin dynamics
$
\dot{s_i}= -F_i({\bf s}) - z(t) s_i + h_i(t)+\eta_i(t),
$
where $\eta$ is a white noise of variance $2T=2/\beta$,
 $z(t)$ is a Lagrange multiplier 
enforcing the constraint and $h_i(t)$ is an external field
(usually set to zero); (ii) the Ising $s_i=\pm 1$
 model with Metropolis dynamics, in which a 
randomly chosen spin flips with probability  $\min [1,\exp(-2 \beta s_i F_i)]$.
The ``energy'' ${\cal E}$ is defined in all cases by the expression
$N{\cal E}({\bf s})=-({\bf F}\cdot{\bf s})/p$ to which only the symmetric part of the 
force contributes.
We monitor the autocorrelation function $C(t,t') \equiv \overline
{\left<{\bf s}(t)\cdot{\bf s}(t')\right>}/N$ and the response
function $G(t,t')\equiv  \sum_i \overline{\delta\left<s_i(t)\right>/\delta h_i(t')}/N$.

Both  models are {\em marginal\/} for $p=2$.
The dynamics of the asymmetric $p=2$ spherical 
and $\pm 1$ (SK)
 models were studied numerically and by mean-field theory (large $N$)
by CS, who found no glassiness at finite $T$ \cite{p2}.
Here we concentrate on the {\em non-marginal\/} case $p>2$.

\begin{figure}
\centerline{\epsfxsize=9cm
\epsffile{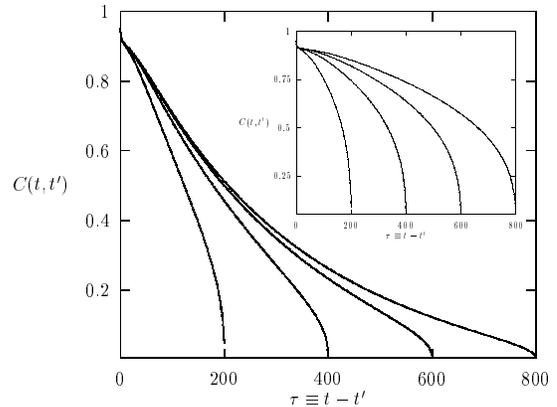}
}

\caption{Interruption of aging: $C(t,t')$ {\it vs.}~$t-t'$
 for $t=200,400,600,800$ at 
           $T=0.2$ and $\alpha=0.96$ compared to aging for $\alpha=1$ (inset).
}

\end{figure}

 We have solved numerically the mean-field equations (exact
for $N \rightarrow \infty$) for the correlation and the response function
of the spherical ($p=3$) model,
starting from a random initial configuration \cite{equa}.
The result for  $\alpha=0.96$ and $T=0.2$ is shown in
Fig.~1.
The corresponding curves for $\alpha=1$ are shown in the inset.
On can see that there is
an initial regime in which the solution exhibits aging
as for $\alpha=1$, followed by a crossover to time-translational
invariance (no aging).  The closer $\alpha$ is to one, 
the longer the (interrupted) aging regime lasts.
We have not found any evidence for a  
critical asymmetry ($\alpha_{\rm c}<1$) beyond 
which the system ages forever although  we cannot rule out a transition 
for $\alpha$ even
closer to one.
The correlations decay to zero ruling out ergodicity-breaking also as in CS.

In order to prove that there are trapping regions for $\alpha <1$, $T > 0$, 
we have studied the mean-field dynamics
starting from a configuration with given (low) energy,
but otherwise random \cite{comment}. This amounts to
solving a {\em static\/} problem for the initial condition, 
at a temperature $T'$ tuned to give the desired value of the
energy \cite{equa2}.
Figure 2 shows  that for $\alpha>0.86$ and sufficiently
small initial energies the system remains trapped at non-zero temperature 
(forever, if $N=\infty$).

\begin{figure}
\centerline{\epsfxsize=9cm
\epsffile{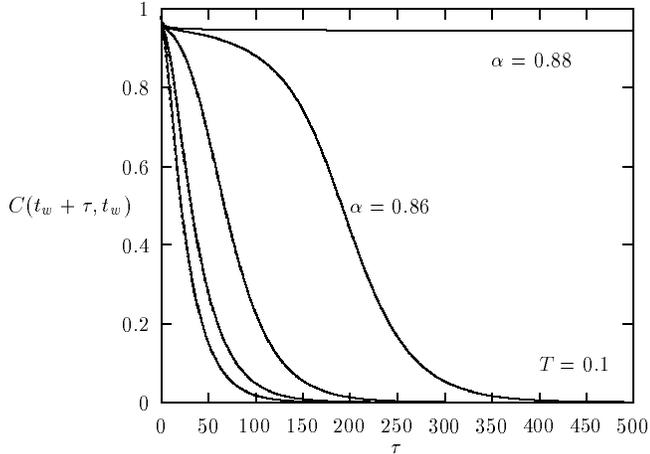}
}

\caption{$C(t_{\rm w}+\tau,t_{\rm w})$
           vs.\ $\tau$ for $t_{\rm w}=100$ at $T=0.1$ starting from an
initial condition with  low energy-density ($T'=0.1$).
From top to bottom
the asymmetry parameter equals $\alpha=0.88,0.86,0.84,0.82,0.8$.
}
\end{figure}

\begin{figure}
\centerline{\epsfxsize=9cm
\epsffile{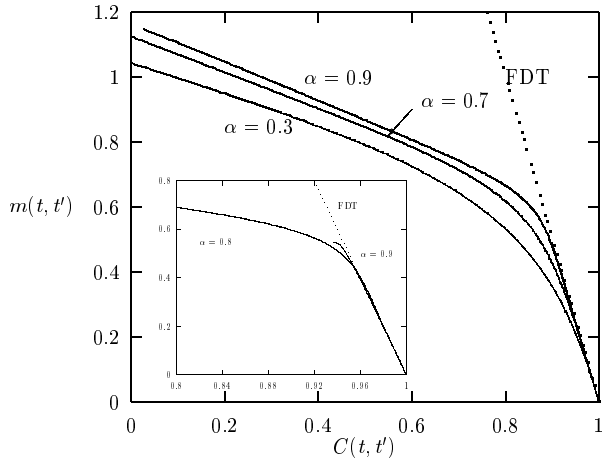}
}

\caption{The magnetization $m(t,t')\equiv \int_0^{t'} ds G(t,s)$
           vs.  $C(t,t')$ starting from a random initial condition 
         for $T=0.2$ and $\alpha=0.3,0.7,0.9$. Inset: same, starting from 
        low energy, $T=0.1$, $\alpha=0.8, 0.9$.
}
\end{figure}

Figure 3 shows the
fluctuation-dissipation (FD) ratio
$G(t,t')=X[C(t,t')](\theta(t-t')/T)\partial C(t,t')/
\partial t'$~\cite{Cuku} starting from random initial conditions. 
The same plot for  $\alpha=1$ 
consists of two straight lines corresponding to $X=1$
for $C>q_{EA}$ (FDT) and $X= {\mbox{constant}} < 1$ for $C<q_{EA}$.
It is remarkable that the plot crosses over smoothly, 
as $\alpha\to 1$, to the one
holding for the relaxational case (note that  $X[C] \leq 1$, $\forall C$, $\forall \alpha$).
In the inset we plot the FD ratio starting from low energy, 
for $\alpha=0.8$ (untrapped) and $\alpha=0.9$ (trapped). 

In conclusion, the mean-field analysis reveals that the  present 
model has trapping regions that cannot be reached when $N=\infty$ because
the time needed to fall into them diverges with $N$ and, by the same
token, that the time needed to {\em escape\/} from a trap also diverges.
The question how typical falling and escaping  times scale with $N$ is beyond the present 
analytical tools. We have thus performed numerical simulations,
choosing for convenience the $\pm 1$ version of the $(p=3)$ model.

Figure~4 shows the typical behaviour at $T=0.01$ and $\alpha=0.5$, for $N=50$.
The system alternates between periods
of trapping and periods of  rapid motion at high energy.
A blow-up in time of the same run shows that 
the overall appearance of the graph is self-similar.
The longest trapping time is of the order of the
total observation time, which indicates a broad distribution
of release times. We have found such behaviour, for $N=50$,
in a region in the  $(T,\alpha)$ plane bounded by 
$\alpha \sim 0.4$ at $T \sim 0$ and $T=T_c \sim 0.05$ at $\alpha=1$.
Traps are visited once showing that there is a large time-span between the 
smallest falling time and the maximal (`equilibration') trapping time.  For larger system 
sizes, $N=100,200$ falling and escaping times increase with $N$,
 as expected.

This behaviour is reminiscent of the non-relaxational dynamics of
a particle in a random velocity field \cite{b1} where 
anomalous diffusion is due \cite{b1,b2} to 
broad, L\'evy-stable distributions of trapping times. Indeed, 
this model  likely to be close to 
a microscopical realisation of the related `trap model' that 
has been fruitfully  used to describe aging in \cite{Traps}.

We have therefore shown that aging and ergodicity breaking resist
non-relaxational perturbations in the dynamics of
disordered systems if they have 
non-marginal states. Finite-size fully connected systems (relevant, e.g., for 
modelling biological networks) may thus  exhibit striking 
aging effects. 
This is also likely
to be the case for finite-dimensional systems whose
mean-field limit is non-marginal. 

Let us also mention that we have checked that 
systems with random forces deriving
from a potential and strongly perturbed by random 
non-potential forces correlated with a different range 
exhibit aging phenomena even at the mean-field level. 

Further scenarios can be envisaged: it is likely that
some no-go results obtained almost a decade ago
did not exhaust all possibilities of nature and
that this direction is open for further research.

\begin{figure}

\centerline{\epsfxsize=9cm
\epsffile{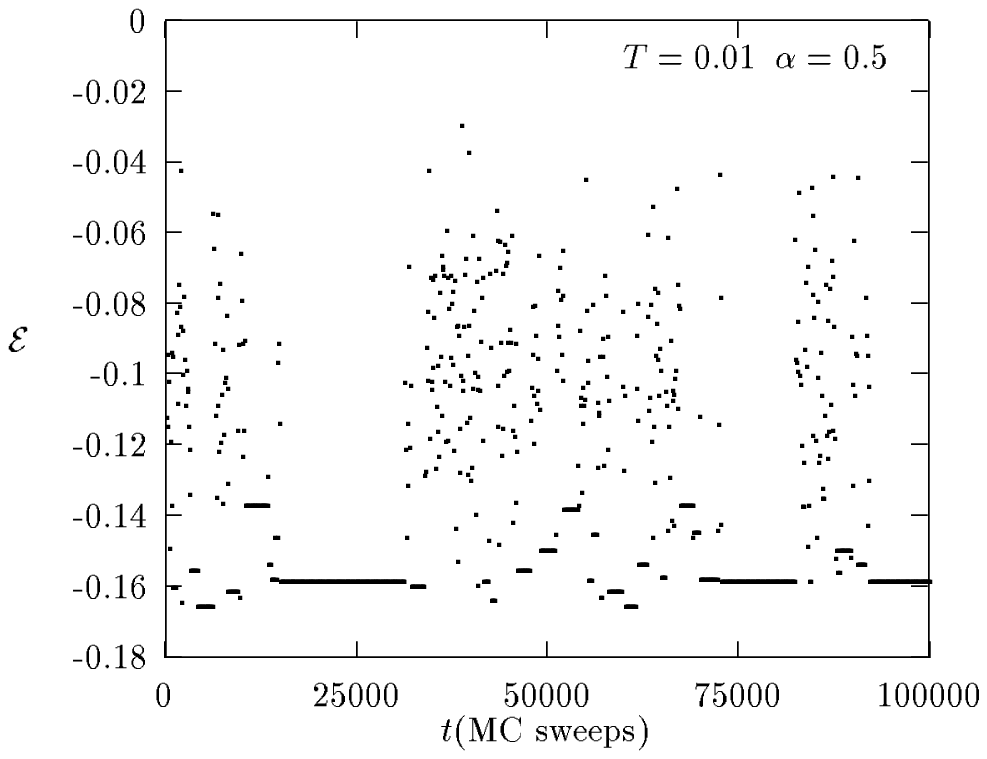}
}

\centerline{\epsfxsize=9cm
\epsffile{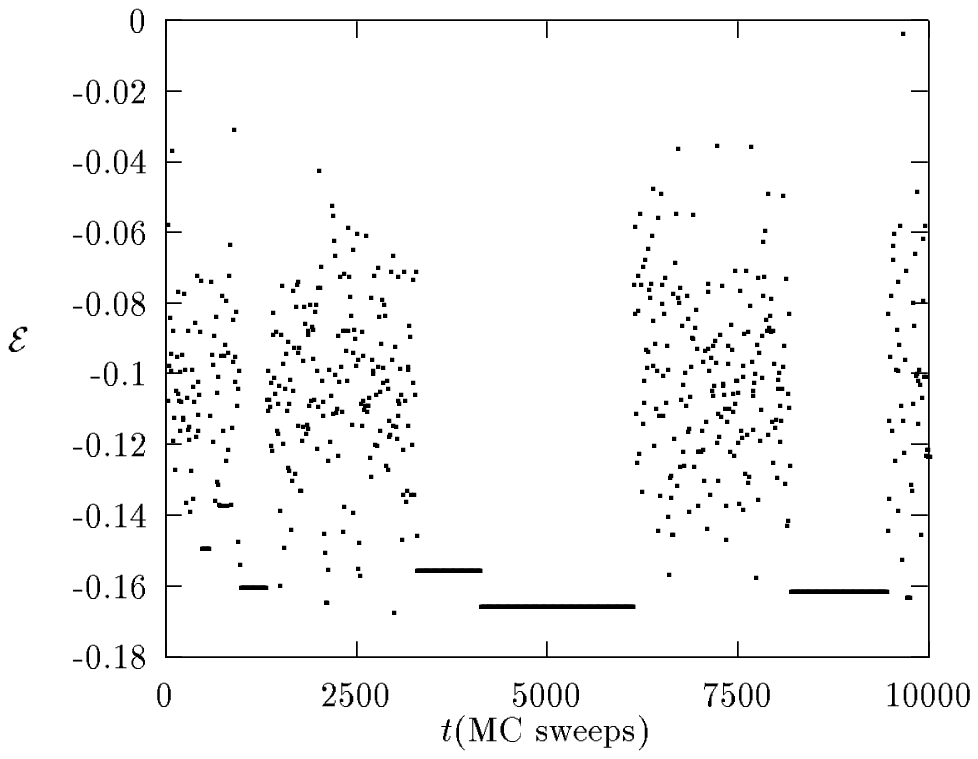}
}

\caption{a. Instantaneous energy ${\cal E}$ vs.\ time
(Monte-Carlo sweeps) in the $\pm 1$
$(p=3)$ model for $\alpha=0.5$ and $T=0.01$. 
b. A blow-up of the first $10\,000$ sweeps.
}

\end{figure}

LP is Associato INFN, Sezione di Napoli, and
is supported by a Chaire Joliot
de l'ESPCI. LFC acknowledges 
support of a CEE contract. 
We thank M. Magnasco for helpful discussions.


\begin{thebibliography}{99}
\bibitem{other}The stability of the ferromagnetic state in the SK model, and of
retrieval states in the Hopfield model had already been pointed out by
\cite{Grheso,CS}.
\bibitem{comment}
Since  the energy involves only the potential part of the  forces, 
small initial energy need not be the best criterion to characterize
trapped situations in highly asymmetric systems.  
We use it here just to show that 
trapping regions exist.
\bibitem{equa2} Again, the method is standard 
\cite{Hojayo,Frpa,Babume}. 
The system starts from the equilibrium configuration of a symmetric system 
with an effective temperature $T_{\rm eff}= T' \sqrt{p/(1+\alpha(p-1))}$. This involves a one step
replica symmetry breaking ansatz with parameters $m$, $q$ and $q_0=0$ 
(as in~\cite{Crso}). The reader may obtain 
the equations governing the subsequent motion from eqs.~(41) of ref.~\cite{Frpa}
 by replacing $f'(q)$ with $f_1(q)$, $f^{\prime\prime}(q)$ with $f_2(q)=\alpha f'_1(q)$, 
and by setting
$\beta'=\sqrt{(1+\alpha(p-1))/p T^2_{\rm eff}}$.
\end{thebibliography}
\end{document}